\documentclass[twocolumn,showpacs,preprintnumbers,amsmath,amssymb]{revtex4}
\usepackage{epsfig}
\usepackage{graphicx}
\usepackage{dcolumn}
\usepackage{bm}

\newcommand{\bea}{\begin{eqnarray}}\newcommand{\eea}{\end{eqnarray}}
\newcommand{\brr}{\begin{array}}\newcommand{\err}{\end{array}}
\newcommand{\bit}{\begin{itemize}}\newcommand{\eit}{\end{itemize}}
\newcommand{\ben}{\begin{enumerate}}\newcommand{\een}{\end{enumerate}}

\def\lab{\label}
\def\lan{\langle}
\def\lf{\left}

\def\non{\nonumber}
\def\ran{\rangle}

\def\ri{\right}

\def\al{\alpha}\def\bt{\beta}

\def\te{\theta}

\def\si{\sigma}
\def\om{\omega}

\def\1{{_{1}}}
\def\2{{_{2}}}

\begin{document}

\title{Neutrino mixing as a source for cosmological constant}

\author{{M.~Blasone${}^{\flat\sharp}$, A.~Capolupo${}^{\flat}$,
S.~Capozziello${}^{\flat}$, S.~Carloni${}^{\natural}$ and
G.~Vitiello${}^{\flat \sharp }$ \vspace{3mm}}}
\email{capolupo@sa.infn.it}

\affiliation{${}^{\flat}$Dipartimento di  Fisica "E.R. Caianiello" and INFN,
Universit\`a di Salerno, I-84100 Salerno, Italy
\\ ${}^{\sharp}$ Unit\`a INFM, Salerno, Italy
\\ ${}^{\natural}$ Department of Applied Mathematics, University of
Cape Town, South Africa}
\date{\today}

\begin{abstract}

We report on recent results showing that neutrino mixing may lead
to a non-zero contribution to the cosmological constant.
This contribution is of a completely different nature with respect to the usual one
by a massive spinor field. We also study the problem of field mixing in Quantum Field Theory
in curved space-time, for the case of a scalar field in the Friedmann-Robertson-Walker metric.
\end{abstract}

\pacs{14.60.Pq; 98.80}

\maketitle

\section{Introduction}

The problem of cosmological constant is currently
one of the most challenging open issues in  theoretical physics and cosmology.
The main difficulty comes from the fact that, when estimating its value as a gravitational
effect of vacuum energy, the numerical results are in strong disagreement with the
accepted  upper bound $\Lambda < 10^{-56} cm^{-2}$
\cite{Zeldovic1}.

We show here that a new contribution to the vacuum energy and therefore to the
cosmological constant may arise from  neutrino mixing \cite{Blasone:2004yh}.
The contribution we find comes from the specific nature of the
field mixing and is therefore of completely different origin with respect to
the ordinary vacuum energy contribution of a massive spinor field.

Indeed, it has been shown \cite{BV95}-\cite{aspects} that
the vacuum for fields with definite masses is not invariant under
the field mixing transformations and in the infinite volume limit,
it is unitarily inequivalent to the vacuum for the fields with
definite flavor.
In the case of neutrinos, this results in a condensate
of neutrino-antineutrino pairs, with a density
related to the mixing angle(s) and mass difference(s) among the
different generations. Phenomenological consequences of such a non-trivial
condensate structure of the flavor vacuum have been studied for neutrino oscillations
\cite{oscill} and for beta decay \cite{nulorentz}.

In this paper we consider the case of two flavors and
Dirac neutrino fields, although the conclusions we reach can be easily
extended to the case of three flavors and Majorana neutrinos \cite{comment,neutral}.

We also include a preliminary study of mixed (bosonic) fields in a curved background,
for the case of FRW metric.

In Section II, we shortly summarize the main
results for neutrino mixing in QFT.
In Section III, we compute  the
neutrino contribution to the  cosmological constant  and
estimate its value by using the natural
scale of neutrino mixing as a cut--off. The result turns out to be
compatible with the currently accepted upper bound on $\Lambda$.
In Section IV, quantum fields and mixing
relations are analyzed in expanding universe. Section V is devoted
to  conclusions.

\section{Neutrino mixing in Quantum Field Theory}

The main features of the QFT formalism for the neutrino mixing are
summarized below. For a detailed review see \cite{aspects}. For sake
of simplicity, we consider the two flavor case and we use Dirac
neutrino fields. The Lagrangian density describing the
Dirac neutrino fields with a mixed mass term is:
\begin{equation}\label{lagemu}{\cal L}(x)\,= \,{\bar \Psi_f}(x) \lf( i
\not\!\partial-M \ri) \Psi_f(x)\  . \end{equation}
The relation
between Dirac fields $\Psi_f(x)$, eigenstates of flavor, and Dirac
fields $\Psi_m(x)$, eigenstates of mass, is given by
\begin{equation}\label{fermix} \Psi_f(x) \, = {\cal U} \, \Psi_m (x).
\end{equation}
${\cal U} $ is the mixing matrix
\begin{equation} \label{fermix1} {\cal U}=\begin{pmatrix}
  \cos\theta & \sin\theta\\
  -\sin\theta & \cos\theta
\end{pmatrix}
\end{equation}
being $\theta$ the
mixing angle. Using
Eq.(\ref{fermix1}), we diagonalize the quadratic form
Eq.(\ref{lagemu}), which then reduces to the Lagrangian for the
Dirac fields $\Psi_m(x)$, with masses $m_i$, $ i =1,2 $ :
\begin{equation}\label{lag12} {\cal L}(x)\,=\,  {\bar \Psi_m}(x) \lf( i
\not\!\partial -  \textsf{M}_d\ri) \Psi_m(x)  \, , \end{equation}
 where
$\textsf{M}_d = diag(m_1,m_2)$.
The mixing transformation
(\ref{fermix}) can be written as \cite{BV95}
\begin{equation}
\label{mt} \nu_{\si}(x)\equiv G^{-1}_{\bf \te}(t) \,
\nu_{i}(x)\, G_{\bf \te}(t), \end{equation}
 where $(\si,i)=(e,1), (\mu,2)$,
and the generator $G_{\bf \te}(t)$ is given by
\begin{equation} G_{\bf \te}(t)=\exp\Big[\te\int d^{3}{\bf
x}\lf(\nu_{1}^{\dag}(x)\nu_{2}(x)-\nu_{2}^{\dag}(x)\nu_{1}(x)\ri)
\Big]  . \end{equation}

The free fields  $\nu_i$ (i=1,2) are given, in the usual way, in
terms of creation and annihilation operators (we use $t\equiv
x_0$):
\begin{equation}\label{2.2} \nu_{i}(x) = \sum_{r} \int \frac{d^3{\bf
k}}{(2\pi)^\frac{3}{2}} \lf[u^{r}_{{\bf k},i}(t) \al^{r}_{{\bf
k},i}\:+    v^{r}_{-{\bf k},i}(t) \bt^{r\dag }_{-{\bf k},i}  \ri]
e^{i {\bf k}\cdot{\bf x}} ,\end{equation}
with $i=1,2$,
\bea u^{r}_{{\bf
k},i}(t)=e^{-i\om_{k,i} t}u^{r}_{{\bf k},i},\qquad v^{r}_{{\bf
k},i}(t)=e^{i\om_{k,i} t}v^{r}_{{\bf k},i}\eea
and
$\om_{k,i}=\sqrt{{\bf k}^2+m_i^2}$.

The mass eigenstate vacuum is denoted by $|0\ran_{m}$:  $\; \;
\al^{r}_{{\bf k},i}|0\ran_{m}= \bt^{r }_{{\bf k},i}|0\ran_{m}=0$.
The anticommutation relations, the wave function orthonormality
and completeness relations are the usual ones (cf. Ref.
\cite{BV95}).

The flavor fields are obtained from Eq. (\ref{mt}):
\bea\non\label{exnue1} \nu_\si(x) & = &\sum_{r} \int
\frac{d^3{\bf k}}{(2\pi)^\frac{3}{2}}
\Big[ u^{r}_{{\bf k},i}(t) \al^{r}_{{\bf k},\si}(t)
\\& + &
v^{r}_{-{\bf k},i}(t) \bt^{r\dag}_{-{\bf k},\si}(t) \Big]
e^{i {\bf k}\cdot{\bf x}} ,
\eea
with $(\si,i)=(e,1) , (\mu,2),$ with the flavor
annihilation operators  defined as
\bea \al^{r}_{{\bf
k},\si}(t) & \equiv & G^{-1}_{\bf \te}(t)\;\al^{r}_{{\bf k},i}\;
G_{\bf \te}(t)\\ \bt^{r}_{{-\bf k},\si}(t) & \equiv &
 G^{-1}_{\bf \te}(t) \;\bt^{r\dag}_{{-\bf k},i}\;
G_{\bf \te}(t).\eea
They annihilate the flavor vacuum $|0(t)\ran_{f}$
given by
\begin{equation}\label{flavac}
|0(t)\ran_{f}\,\equiv\,G_{\te}^{-1}(t)\;|0\ran_{m} \;. \end{equation}

In the infinite volume limit, the
vacuum $|0(t)\ran_{f}$ for the flavor fields and the vacuum
$|0\ran_{m}$ for the fields with definite masses are unitarily
inequivalent vacua \cite{BV95,hannabuss}.

One further remark is that the use of the vacuum state $|0
{\rangle}_{m}$ in the computation of the two point Green's
functions leads to the violation of the probability conservation
\cite{oscill}. The correct result is instead obtained by the use
of the flavor vacuum $|0 {\rangle}_{f}$, which is therefore the
relevant vacuum to be used in the computation of the oscillation
effects. We will thus use $|0 {\rangle}_{f}$ in our computations
in the following.

The explicit
expressions for the flavor annihilation/creation operators in the
reference frame ${\bf k}=(0,0,|{\bf k}|)$ are \cite{BV95}:
\bea \non \al_{{\bf k},e}^{r}(t)&=&\cos\theta \al_{{\bf k},1}^{r} +
\sin\theta\lf(U_{\bf k}^*(t) \al_{{\bf k},2}^{r} +\epsilon^{r}
V_{\bf k}(t) \bt_{-{\bf k},2}^{r\dag}\ri)
\\ \non\al_{{\bf k},\mu}^{r}(t)&=& \cos\theta \al_{{\bf k},2}^{r} -
\sin\theta\lf(U_{\bf k}(t) \al_{{\bf k},1}^{r} - \epsilon^{r}
V_{\bf k}(t) \bt_{-{\bf k},1}^{r\dag}\ri)
\\\non \bt^{r}_{-{\bf k},e}(t)&=&\cos\theta \bt_{-{\bf k},1}^{r} + \sin
\theta \lf(U_{\bf k}^*(t) \bt_{-{\bf k},2}^{r}
-\epsilon^{r}V_{\bf k}(t) \al_{{\bf k},2}^{r\dag}\ri)
\\ \non\bt^{r}_{-{\bf k},\mu}(t)&=&\cos\theta \bt_{-{\bf k},2}^{r} -\sin
\theta \lf(U_{\bf k}(t) \bt_{-{\bf k},1}^{r} +\epsilon^{r}V_{\bf
k}(t) \al_{{\bf k},1}^{r\dag}\ri)
\\ \eea
where $U_{\bf k}$ and $V_{\bf k}$  are  Bogoliubov coefficients given
by:
\bea V_{\bf k}(t)&=&|V_{\bf
k}|\;e^{i(\om_{k,2}+\om_{k,1})t},
\\ U_{\bf k}(t)&=&|U_{\bf k}|\;e^{i(\om_{k,2}-\om_{k,1})t}, \eea
\bea
\non |U_{\bf k}|&=&\lf(\frac{\om_{k,1}+m_{1}}{2\om_{k,1}}\ri)
^{\frac{1}{2}}
\lf(\frac{\om_{k,2}+m_{2}}{2\om_{k,2}}\ri)^{\frac{1}{2}}
\\\non& \times &\lf(1+\frac{|{\bf k}|^{2}}{(\om_{k,1}+m_{1})
(\om_{k,2}+m_{2})}\ri) ,
\\ [2mm] \non \label{V}
|V_{\bf k}|&=&\lf(\frac{\om_{k,1}+m_{1}}{2\om_{k,1}}\ri)
^{\frac{1}{2}}
\lf(\frac{\om_{k,2}+m_{2}}{2\om_{k,2}}\ri)^{\frac{1}{2}}
\\ & \times &\lf(\frac{|{\bf k}|}{(\om_{k,2}+m_{2})}-\frac{|{\bf
k}|}{(\om_{k,1}+m_{1})}\ri) , \eea
with
\begin{equation} |U_{\bf k}|^{2}+|V_{\bf k}|^{2}=1. \end{equation}

The function $|V_{\bf k}|$ is related to the condensate content of
the flavor vacuum \cite{BV95} as:
\bea\label{con}
{}_{e,\mu}\langle 0| \al_{{\bf k},i}^{r \dag} \al^r_{{\bf k},i}
|0\rangle_{e,\mu}\,= \, \sin^{2}\te\; |V_{{\bf
k}}|^{2}  \;, \quad  i=1,2\eea
with the same result for antiparticles.

\begin{figure}
\begin{center}
\includegraphics*[width=8.5cm]{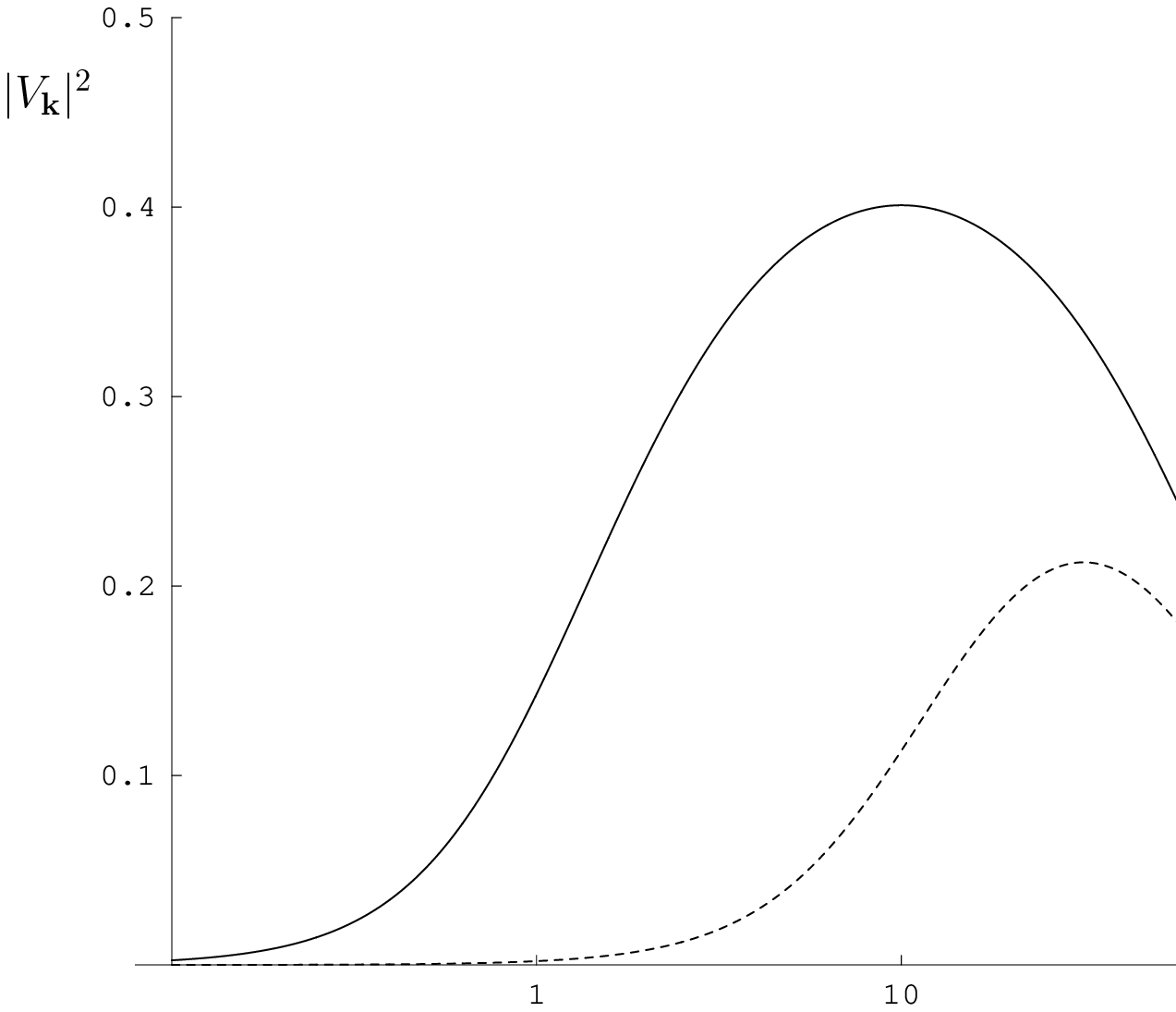}
\vspace{-.5cm}
\end{center}
\caption{Fermion condensation
density $|V_{\bf k}|^2$ as a function of  ${|\bf k|}$
and for sample values of the parameters $m_1$ and $m_2$.
Solid line: $m_1=1$, $m_2=100$; Long dashed line: $m_1=10$, $m_2=100$; Short dashed
line: $m_1=10$, $m_2=1000$.} \label{fig1} \end{figure}
\medskip

In Fig.1, we plot the fermion condensation density $|V_{\bf k}|^2$ in
function of  ${|\bf k|}$ and for sample values of the parameters
$m_1$ and $m_2$.
$|V_{\bf k}|^2$ is zero for $m_{1} =
m_{2}$, it has a maximum at $|{\bf k}|=\sqrt{m_\1 m_\2}$ and, for
$ |{\bf k}| \gg\sqrt{m_\1 m_\2}$, it goes like $|V_{{\bf
k}}|^2\simeq (m_\2 -m_\1)^2/(4 |{\bf k}|^2)$.

From the study of the current algebra, we obtain the flavor charge
operators that, in terms of flavor operators, are:
\bea\label{charneut} Q_{\sigma }(t)=\sum_{{\bf k},r}\left( \alpha
_{{\bf k},\sigma }^{r\dagger }(t)\alpha _{{\bf k},\sigma
}^{r}(t)-\beta _{-{\bf k} ,\sigma }^{r\dagger }(t)\beta _{-{\bf
k},\sigma }^{r}(t)\right) , \eea
with $\sigma =e,\mu.$

 At time $t=0$, the vacuum state is $|0\rangle_{e,\mu}$ and the one
electron neutrino state is defined as:
\bea\non |\nu_e \rangle \equiv \alpha_{{\bf k},e}^{r
\dag}|0\rangle_{e,\mu} \label{h1}\eea

Considering the
flavor charge operators, defined as in Eq.(\ref{charneut}). We
then have (in the Heisenberg representation) \bea
 _{e,\mu }\langle 0|Q_{e}(t)|0\rangle _{e,\mu }=_{e,\mu
}\langle 0|Q_{\mu }(t)|0\rangle _{e,\mu }=0, \eea
\bea
{\cal Q}^e_{{\bf k},e}(t)&=&\langle \nu _{e}|Q_{e}(t)|\nu _{e}\rangle
\\\non &=&\lf|\lf \{\al^{r}_{{\bf k},e}(t), \al^{r \dag}_{{\bf k},e}(0)
\ri\}\ri|^{2} + \lf|\lf\{\bt_{{-\bf k},e}^{r \dag}(t), \al^{r
\dag}_{{\bf k},e}(0) \ri\}\ri|^{2},\eea
\bea {\cal Q}^e_{{\bf
k},\mu}(t) &=&\langle \nu _{e}|Q_{\mu }(t)|\nu _{e}\rangle
\\\non &=& \lf|\lf
\{\al^{r}_{{\bf k},\mu}(t), \al^{r \dag}_{{\bf k},e}(0)
\ri\}\ri|^{2}+ \lf|\lf\{\bt_{{-\bf k},\mu}^{r \dag}(t),
\al^{r \dag}_{{\bf k},e}(0) \ri\}\ri|^{2}. \eea

Charge conservation is obviously ensured at any time: \bea {\cal
Q}^e_{{\bf k},e}(t)+{\cal Q}^e_{{\bf k},\mu}(t)=1. \eea

The oscillation formula for the flavor charges are then
\cite{oscill}:
\bea\non\label{oscillfor1}
{\cal Q}^e_{{\bf k},e}(t)&=& 1-\sin ^{2}(2\theta
)\Big[ \left| U_{\mathbf{k}}\right| ^{2}\sin
^{2}\left( \frac{\omega _{k,2}-\omega _{k,1}}{2}t\right)
\\&+& \left| V_{\mathbf{k%
}}\right| ^{2}\sin ^{2}\left( \frac{\omega _{k,2}+\omega
_{k,1}}{2}t\right) \Big] ,
\eea
\bea\non\label{oscillfor2}
{\cal Q}^e_{{\bf k},\mu}(t) &=&\sin ^{2}(2\theta )\Big[
\left| U_{\mathbf{k}}\right| ^{2}\sin
^{2}\left( \frac{\omega _{k,2}-\omega _{k,1}}{2}t\right)
\\&+&\left| V_{\mathbf{k%
}}\right| ^{2}\sin ^{2}\left( \frac{\omega _{k,2}+\omega
_{k,1}}{2}t\right) \Big] .
\eea

This result is exact. There are two differences with respect to
the usual formula for neutrino oscillations: the amplitudes are
energy dependent, and there is an additional oscillating term.

For $\left| \mathbf{k}\right| \gg \sqrt{m_{1}m_{2}}$ we have
$\left| U_{\mathbf{k}}\right| ^{2}\longrightarrow 1$ and $\left|
V_{\mathbf{k}}\right| ^{2}\longrightarrow 0$ and the traditional
formula is recovered.

Similar results are obtained in the case of boson fields \cite{bosonmix}.
For the $\eta-\eta'$
system, the correction may be as large as
$20\%$.

\section{Neutrino mixing contribution to the cosmological
constant}

The connection between the vacuum energy density
$\lan\rho_{vac}\ran$ and the cosmological constant $\Lambda$ is
provided by the well known relation
\bea\label{ed} \lan\rho_{vac}\ran= \frac{\Lambda}{4\pi G} , \eea
where $G$ is the gravitational constant.

 The energy--momentum tensor density
${\cal T}_{\mu \nu}$ is obtained  by varying the action with
respect to the metric $g_{\mu\nu}$:
 \bea\ \label{impulsoenergia}
{\cal T}_{\mu\nu}=\frac{2}{\sqrt{-g}}\frac{\delta S}{\delta
g^{\mu\nu}(x)} ,
 \eea
where the action is
 \bea\
S=\int \sqrt{-g}{\cal L}(x) d^{4}x.
 \eea%

In the present case, the energy momentum tensor density is given
by
\bea\label{tmn}
 {\cal T}_{\mu\nu}(x) = \frac{i}{2}\left({\bar
\Psi}_{m}(x)\gamma_{\mu}(x) \overleftrightarrow{D}_{\nu}
\Psi_{m}(x)\right)\ ,
 \eea
where $\overleftrightarrow{D}_{\nu}$ is the covariant derivative:
\begin{equation}\label{der spin}
D_{\nu}=\partial_{\nu}+\Gamma_{\nu},
 \quad\; \Gamma_{\nu}=\frac{1}{8} \omega^{a
b}_{\nu}[\gamma_{a},\gamma_{b}], \quad\;
  \gamma_{\mu}(x)=\gamma^{c}e_{c \mu}(x),
\end{equation}
 being $\gamma^{c}$ the standard Dirac matrices, and
 $\bar{\Psi}\overleftrightarrow{D}_{\nu}\Psi=\bar{\Psi}D\Psi-(D\bar{\Psi})
\Psi$. Let us consider the Minkowski metric, we have
\bea\
 {\cal T}_{00}(x) = \frac{i}{2}:\left({\bar \Psi}_{m}(x)\gamma_{0}
\overleftrightarrow{\partial}_{0} \Psi_{m}(x)\right):\eea where
$:...:$ denotes  the customary normal ordering with respect to the
mass vacuum in the flat space-time.

In terms of the annihilation and creation operators of fields
$\nu_{1}$ and $\nu_{2}$, the energy-momentum tensor \bea
T_{00}=\int d^{3}x {\cal T}_{00}(x)\eea is given by
\bea T^{(i)}_{00}= \sum_{r}\int d^{3}{\bf k}\,
\omega_{k,i}\lf(\al_{{\bf k},i}^{r\dag} \al_{{\bf k},i}^{r}+
\beta_{{\bf -k},i}^{r\dag}\beta_{{\bf
-k},i}^{r}\ri),\eea
with $i=1,2$.

Note that $T^{(i)}_{00}$ is time independent.

The expectation value of
$T^{(i)}_{00}$ in the flavor vacuum $| 0\ran_f$ gives the contribution
 $\lan\rho_{vac}^{mix}\ran$ of the neutrino mixing
to the vacuum energy density is:
 \bea\
 {}_f\lan 0 |\sum_{i} T^{(i)}_{00}(0)|
0\ran_f = \lan\rho_{vac}^{mix}\ran \eta_{00} ~.
 \eea

Within the QFT formalism for neutrino mixing, we have
 \bea {}_f\lan 0 |T^{(i)}_{00}| 0\ran_f={}_f\lan
0(t) |T^{(i)}_{00}| 0(t)\ran_f \eea for any t. We then obtain
\bea \non {}_f\lan 0 |\sum_{i} T^{(i)}_{00}(0)| 0\ran_f &=&
\sum_{i,r}\int d^{3}{\bf k} \, \omega_{k,i}\Big({}_f\lan 0
|\al_{{\bf k},i}^{r\dag} \al_{{\bf k},i}^{r}| 0\ran_f
\\&+& {}_f\lan 0
|\beta_{{\bf k},i}^{r\dag} \beta_{{\bf k},i}^{r}| 0\ran_f \Big) .
\eea
By using Eq.(\ref{con}),
we get
\bea\non\label{aspT} {}_f\lan 0 |\sum_{i} T^{(i)}_{00}(0)| 0\ran_f
&=&\,8\sin^{2}\theta \int d^{3}{\bf
k}\lf(\omega_{k,1}+\omega_{k,2}\ri) |V_{\bf k}|^{2}
\\&=&\lan\rho_{vac}^{mix}\ran \eta_{00}, \eea
i.e.
\bea\label{cc} \lan\rho_{vac}^{mix}\ran = 32 \pi^{2}\sin^{2}\theta
\int_{0}^{K} dk \, k^{2}(\omega_{k,1}+\omega_{k,2}) |V_{\bf
k}|^{2} , \eea
where the cut-off $K$ has been introduced. Eq.(\ref{cc}) is our
result: it shows that the cosmological constant gets a non-zero
contribution induced from the neutrino mixing \cite{Blasone:2004yh}.
Notice that such a contribution is indeed zero in the no-mixing
limit when the mixing angle $\theta = 0$ and/or $m_{1} = m_{2}$.
Moreover, the contribution is absent in the traditional
phenomenological (Pontecorvo) mixing treatment.

We may try to estimate the neutrino mixing contribution by making
our choice for the cut-off. If we choose the cut-off proportional
to the natural scale appearing in the mixing phenomenon
$\textbf{k}_{0}\simeq \sqrt{m_{1} m_{2}}$  \cite{BV95}: using
$K\sim k_{0}$, $m_{1}=7 \times 10^{-3}eV$, $m_{2}=5 \times
10^{-2}eV$,
 and
 $\sin^{2}\theta\simeq 0.3$ \cite{masses} in Eq.(\ref{cc}), we obtain
\bea \lan\rho_{vac}^{mix}\ran =0.43 \times 10^{-47}GeV^{4}\eea

Using Eq.(\ref{ed}), we are in agreement with the upper bound
 for $\Lambda$:
\bea \Lambda \sim 10^{-56}cm^{-2} , \eea
Another possible choice is to use the electro-weak scale cut-off:
$K\approx 100 GeV$. We then have \bea \lan\rho_{vac}^{mix}\ran
=1.5 \times 10^{-15}GeV^{4}\eea and \bea \Lambda \sim
10^{-24}cm^{-2} , \eea which is, however, beyond the accepted
upper bound.

In a recent paper \cite{Barenboim:2004ev}, it was suggested the cut-off
scale given by the sum of the two neutrino masses,
$K = m_{1} + m_{2}$.

For hierarchical neutrino models, for which ${m_{2}\gg m_{1}}$, we
have, in this case, $ K \gg \sqrt{m_{1} m_{2}},$ and thus, if we
assume that the modes near the cut--off contribute mainly to the
vacuum energy,
 and take into account the asymptotic properties of $V_{{\bf k}}$:
\bea|V_{{\bf k}}|^2 \simeq \frac{(m_2 -m_1)^2}{4 k^2} \qquad\qquad
 k\gg\sqrt{m_1 m_2}\eea
 we obtain:
\bea
\non\lan\rho_{vac}^{mix} \ran & \sim &  8\pi \sin^{2}\theta(m_2 -m_1)^{2}
(m_\2 +m_\1)^{2}\times
\\
& \times &\lf(\sqrt{2}+1+O \lf(\frac{m^{2}_1}{m^{2}_2} \ri) \ri)
\eea
and then
\bea \lan\rho_{vac}^{mix}\ran \propto \sin^{2}\theta
(\Delta m^{2})^{2}\eea
in the limit
$m_{2}\gg m_{1}$.

In Ref.\cite{Barenboim:2004ev}, the corresponding $\Delta m^{2}$
is given by the solar neutrino data:
 $\Delta m^{2}\simeq  10^{-5}eV^{2}$, resulting
 in a contribution of the right order.

\section{Quantum fields and mixing in expanding universe}

In this Section we present a preliminary study of mixed fields
on a time--dependent gravitational background. For simplicity, we
consider the case of  neutral scalar fields, the case of fermionic fields
and neutrino oscillations will  be considered elsewhere.

\subsection{Free fields in expanding universe}

We start by quantizing a
free neutral scalar field $\phi$ in the Friedmann--Robertson--Walker (FRW)
space-time with flat spatial sections, characterized by metrics of
the form
\bea
ds^{2}=g_{\mu\nu}dx^{\mu}dx^{\nu}=dt^{2}-a^{2}(t)d{\bf x}^{2},
\eea
where $a(t)$ is the scale factor.

A flat FRW space-time is a conformally flat space-time. Indeed, by
replacing the coordinate $t$ by the conformal time $\eta$,
\bea
\eta(t)=\int_{t_{0}}^{t}\frac{dt}{a(t)},
\eea
where $t_{0}$ is an arbitrary constant, the line element takes the form
\bea
ds^{2}=a^{2}(\eta)[d\eta^{2}-d{\bf x}^{2}],
 \eea
where $a(\eta)$
is the scale factor expressed through the new variable $\eta$.
Introducing the auxiliary field $\chi=a(\eta)\phi$, it is possible
to show that the evolution of a scalar field $\phi$ in a flat FRW
metric is mathematically equivalent to the dynamics of the
auxiliary field $\chi$ in the Minkowski metric \cite{Mukhanov}. The
information about the influence of gravitational field on  $\phi$
is contained in the time-dependent mass $m_{e\!f\!f}(\eta)$ defined by
\bea m_{e\!f\!f}^{2}(\eta)=m^{2}a^{2}-\frac{a''}{a},\eea where the
prime $'$ denotes the derivative with respect to $\eta$.

The field $\chi$ can be quantized in the standard fashion by
introducing the equal time commutation relations
\bea [\chi({\bf
x},\eta), \pi({\bf y},\eta)]=i \delta^3(\bf x - \bf y), \eea
where
$\pi=\chi'$ is the canonical momentum.

The Hamiltonian of the
quantum field $\chi$ is
\bea\label{Hamcurved}
H(\eta)=\frac{1}{2}\int d^{3} {\bf x} \lf[\pi^{2}+(\nabla
\chi)^{2}+ m_{e\!f\!f}^{2}(\eta)\chi^{2} \ri]. \eea
Note that the
energy of the field $\chi$ is not conserved; this leads to the
possibility of particle creation in the vacuum. The energy for new
particles is supplied by gravitational field.

The field $\chi$ is expanded as \bea\label{chimass} \chi ({\bf
x},\eta) = \int \frac{d^3{\bf k}}{(2\pi)^{\frac{3}{2}}} \frac{e^{i
{\bf k x}}}{\sqrt{2}} \lf(v_{k}^{*}(\eta) a_{\bf k} + v_{k}(\eta)
a^{{\dag}}_{-{\bf k}} \ri), \eea where the mode functions
$v_{k}(\eta)$ obey the equations \bea\label{modefunctions} v_{k}''
+\omega_{k}^{2}(\eta)v_{k}=0, \qquad \omega_{k}(\eta) =
\sqrt{k^{2}+m_{e\!f\!f}^{2}(\eta)}\eea and satisfy the following
normalization condition \bea v'_{k}v^{*}_{k}-
v_{k}v'^{*}_{k}=2i.\eea
We have $v_{k}(\eta) = v_{-k}(\eta)$ as follows
from the relation $(\chi_{\bf k})^{*}=\chi_{-\bf k}$
and $(a_{\bf k})^{*}=a^{\dag}_{\bf k}$, where
\bea
\chi_{\bf k}(\eta)=\frac{1}{\sqrt{2}} \lf(v_{k}^{*}(\eta) a_{\bf k} +
v_{k}(\eta)
a^{{\dag}}_{-{\bf k}} \ri).
\eea

The annihilation operators are given in terms of $\chi_{\bf
k}(\eta)$ and $v_{k}(\eta)$ by: \bea a_{\bf k}=\sqrt{2}\;
\frac{v'_{k}\chi_{\bf k}-v_{k}\chi'_{\bf
k}}{v'_{k}v^{*}_{k}-v_{k}v'^{*}_{k}}. \eea Note that $a_{\bf k}$
is time--independent; the commutation relations are: \bea [a_{\bf
k},a^{\dag}_{\bf k'}]=\delta^3 ({\bf k}-{\bf k'}), \qquad [a_{\bf
k},a_{\bf k'}]=[a^{\dag}_{\bf k},a^{\dag}_{\bf k'}]=0.\eea

The vacuum state annihilated by $a_{\bf k}$ is denoted by
$|0(\eta)\rangle$: $a_{\bf k}|0(\eta)\rangle=0$.

\subsection{The instantaneous vacuum}

The Hamiltonian Eq.(\ref{Hamcurved}) is time dependent, then we can
define an ``instantaneous'' vacuum by selecting an arbitrary time
$\eta_{0}$ and defining the vacuum $|0(\eta_{0})\rangle$ as the
lowest energy eigenstate of the Hamiltonian $H(\eta_{0})$ computed
at the time $\eta_{0}$.

 The mode functions $v_{k}(\eta_{0})$,
corresponding to the vacuum $|0(\eta_{0})\rangle$, are obtained by
computing the expectation value $\langle
0(\eta)|H(\eta_{0})|0(\eta)\rangle$  in the vacuum state
$|0(\eta)\rangle$ determined by arbitrary chosen mode functions
$v_{k}(\eta)$ and then minimizing that expectation value with
respect to all possible choice of $v_{k}(\eta)$ \cite{Mukhanov}.

The Hamiltonian Eq.(\ref{Hamcurved}) expressed through the
annihilations operators $a_{\bf k}$ defined by the arbitrary mode
functions $v_{k}(\eta)$
is
\bea \non H(\eta)&=&\frac{1}{4}\int
d^{3}{\bf k}
\Big[\Big(v'^{2}_{k}(\eta)+\omega_{k}^{2}(\eta)v^{2}_{k}(\eta)\Big)
^{*}a_{\bf k}a_{-\bf k}
\\\non &+&
\Big(v'^{2}_{k}(\eta)+\omega_{k}^{2}(\eta)v^{2}_{k}(\eta)\Big)
a^{\dag}_{\bf k}a^{\dag}_{-\bf k}
\\\non &+&
\Big(|v_{k}'(\eta)|^{2}+\omega_{k}^{2}(\eta)|v_{k}(\eta)|^{2}\Big)
\Big(2 a^{\dag}_{\bf k}a_{\bf k}+\delta^{3}(0)\Big)\Big].
\\\eea

Since $a_{\bf k}|0(\eta)\rangle=0$, we have
\bea\non \langle
0(\eta)|H(\eta_{0})|0(\eta)\rangle &=& \frac{1}{4}\delta^{3}(0)
\int d^{3}{\bf k}\Big(|v_{k}'(\eta)|^{2}
\\\non &+&\omega_{k}^{2}(\eta)|v_{k}(\eta)|^{2} \Big)_{\eta=\eta_{0}} \eea
and the density energy is
\bea \rho = \frac{1}{4}\int d^{3}{\bf
k}\Big(|v_{k}'(\eta)|^{2} +\omega_{k}^{2}(\eta)|v_{k}(\eta)|^{2}
\Big)_{\eta=\eta_{0}}. \eea

At fixed value of the momentum $k$, if
$\omega_{k}^{2}(\eta_{0})>0$, it is possible to show that the mode
functions $v_{k}(\eta)$ that minimize $\rho$ satisfy the following
initial conditions at $\eta=\eta_{0}$ \cite{Mukhanov}:
\bea\label{condiniz}
v_{k}(\eta_{0})=\frac{1}{\sqrt{\omega_{k}(\eta_{0})}},\qquad
v'_{k}(\eta_{0})= i \sqrt{\omega_{k}(\eta_{0}}).\eea

The mode functions satisfying the Eq.(\ref{condiniz}) define the
annihilation operators $a^{0}_{\bf k}$ of the vacuum
$|0(\eta_{0})\rangle$: $a^{0}_{\bf k}|0(\eta_{0})\rangle=0$
through which the Hamiltonian $H(\eta_{0})$ at time $\eta_{0}$ is
expressed as
\bea H(\eta_{0})= \int d^{3}{\bf k} \,
\omega_{k}(\eta_{0})\Big( a^{0\dag}_{\bf k}a^{0}_{\bf
k}+\frac{1}{2}\delta^{3}(0)\Big).\eea

The zero point energy density of quantum field in the vacuum state
$|0(\eta_{0})\rangle$ is
\bea \rho_{0} = \frac{1}{2}\int d^{3}{\bf
k}\;\omega_{k}(\eta_{0}). \eea
This quantity is time dependent, but,
considering the problem of particle oscillations in the present
time, since the time scale of mixing phenomena are much smaller than
the cosmological time scale, we can neglect the particle creation
in the vacuum due to the gravitational field and we may consider
$\rho \simeq constant$ at the present time and then we renormalize in
usual way.

For further analysis of the vacuum structure in a curved
 background see refs.\cite{Martellini:1978sm}.

\subsection{Mixed fields in expanding universe}

The boson mixing relations in FRW space-time are generalized as
\bea \non
&&\chi_{A}({\bf x},\eta) = \chi_{1}({\bf x},\eta) \; \cos\te + \chi_{2}({\bf x},\eta) \; \sin\te
\\[2mm] \lab{mixcurved}
&&\chi_{B}({\bf x},\eta) =- \chi_{1}({\bf x},\eta)\; \sin\te + \chi_{2}({\bf x},\eta)\; \cos\te
\eea

We now proceed in a similar way to what has been done in
Ref.\cite{bosonmix} for bosons in flat space-time
and recast Eqs.(\ref{mixcurved}) into the form:
\bea\non
\chi_{A}({\bf x},\eta) = G^{-1}_\te(\eta)\; \chi_{1}({\bf x},\eta)\;
 G_\te(\eta) \\[2mm]
\lab{Gmixcurved}
\chi_{B}({\bf x},\eta) = G^{-1}_\te(\eta)\; \chi_{2}({\bf x},\eta)\;
G_\te(\eta)
\eea
and similar ones for $\pi_{A}({\bf x},\eta)$, $\pi_B({\bf x},\eta)$.
$G_\te(\eta)$  denotes the
operator which implements the mixing transformations (\ref{mixcurved}):
\bea
G_\te(\eta) = \exp[\te S(\eta) ] ,
\eea
with \bea\non
S(\eta)&=& -i\; \int d^{3}{\bf x}\,
\Big(\pi_{1}({\bf x},\eta)\chi_{2}({\bf x},\eta)
\\
&-&\pi_{2}({\bf x},\eta)\chi_{1}({\bf x},\eta) \Big).
\eea
We have, explicitly
\bea\non
S(\eta)&=& \int d^3 {\bf k}\,\Big( U^*_{k}(\eta) \, a_{{\bf
k},1}^{\dag}a_{{\bf k},2} -
V_{k}^{*}(\eta) \, a_{-{\bf k},1}a_{{\bf k},2}
\\
 &+& V_{k}(\eta) \,
a_{-{\bf k},1}^{\dag}a_{{\bf k},2}^{\dag}
- U_{k}(\eta) \, a_{{\bf k},1}a_{{\bf k},2}^{\dag} \Big)
\eea
where
 $U_{k}(\eta)$ and $V_{k}(\eta)$
are Bogoliubov coefficients given by
\bea
U_{k}(\eta) &\equiv& i
\lf[ v'^{*}_{k,1}(\eta) v_{k,2}(\eta) - v'_{k,2}(\eta)v_{k,1}^{*}(\eta)
\ri] ,
\\
V_{k}(\eta) &\equiv& -i
\lf[ v'_{k,1}(\eta) v_{k,2}(\eta) - v'_{k,2}(\eta) v_{k,1}(\eta)
\ri].  \eea
and satisfy the relation
\bea\
&&|U_{k}|^{2}-|V_{k}|^{2}=1\,.
\eea
Similar results can be obtained in the case of fermion mixing.

\subsection{The de Sitter space-time}

The de Sitter space-time is characterized by a scale factor of the
form \bea a(t) \sim e^{Ht}, \eea where $H=\dot{a}/a>0$ is the
Hubble constant. The conformal time $\eta$, the scale factor
$a(\eta)$ and the effective frequency are given by \bea \eta = -
\frac{1}{H}e^{-Ht}, \qquad a(\eta)=-\frac{1}{H \eta},
\\
\omega_{k}^{2}(\eta)=k^{2}+\Big(\frac{m^{2}}{H^{2}}-2
\Big)\frac{1}{\eta^{2}}.\eea The conformal time $\eta$ ranges from
$-\infty$ to $0$ when the proper time $t$ goes from $-\infty$ to $
\infty$. The origin of $\eta$ is chosen so that the infinite
future corresponds to $\eta=0$.

For this space-time, the general solution of Eq.(\ref{modefunctions}) is
\bea
v_{k}(\eta)=\sqrt{k|\eta|}\,\Big[A J_{n}(k|\eta|)+ B Y_{n}(k|\eta|)\Big],
\eea
with $n=\sqrt{\frac{9}{4}-\frac{m^{2}}{H^{2}}}$, A and B
constants, and $J_{n}(k|\eta|)$ and $Y_{n}(k|\eta|)$ Bessel functions.

In the de Sitter space-time, a suitable vacuum state is the
Bunch-Davies vacuum, defined as the Minkowski vacuum in the early
time limit ($\eta \rightarrow -\infty$):
\bea v_{k}(\eta)=
\frac{1}{\sqrt{\omega_{k}}}e^{i \omega_{k} \eta },  \qquad \eta
\rightarrow -\infty. \eea
In this case, the mode functions are
\bea v_{k}(\eta)=\sqrt{\frac{\pi|\eta|}{2}}\Big[J_{n}(k|\eta|)- i
Y_{n}(k|\eta|)\Big], \eea
with
$n=\sqrt{\frac{9}{4}-\frac{m^{2}}{H^{2}}}$.

Assuming for the present time, i.e. the actual time relative to
the observer,
\bea
t = 0 \; \Rightarrow \; \eta = -\frac{1}{H}
\; \Rightarrow \; a(\eta)=1,
\eea
the mixing transformations in the flat
space-time are good approximation of those in FRW space-time
Eqs.(\ref{mixcurved}), since the time scale of mixing phenomena
are much smaller than the cosmological time scale. The results
obtained in Section III describe the contribution to the value of
the cosmological constant given by the neutrino mixing at present
time.

\section{Conclusions}

The neutrino mixing is a possible source for the cosmological
constant. Indeed, the non--perturbative vacuum structure
associated with neutrino mixing leads to a non--zero contribution
to the value of the cosmological constant \cite{Blasone:2004yh}.
The value of $\Lambda$ consistent with its accepted upper
bound is found by using the natural scale of the neutrino mixing
cut--off.
The origin of the contribution here discussed is completely different
from that of the ordinary contribution to the vacuum zero energy
of a massive spinor field: it is the {\it mixing}
phenomenon which provides the  vacuum energy contribution discussed
in this paper.

\section{Acknowledgments}

We thank MURST, INFN, INFM and ESF Program COSLAB for financial support.

\bibliography{apssamp}

\end{document}